\documentclass[sigplan,screen]{acmart}
\usepackage{iris}
\usepackage{mathpartir}
\usepackage{cleveref}

\makeatletter

\newcommand*{\fforallAuxx}[1]{%
	\sbox0{$\m@th#1\forall$}%
	\sbox2{%
		\rlap{%
			\raisebox{\depth}{$\m@th#1\backslash$}%
		}%
		\kern\ht0 %
	}%
	\sbox2{\resizebox{\ht2}{\height}{\copy2}}%
	\sbox2{\resizebox{!}{\ht0}{\copy2}}%
	\wd2=0pt %
	\copy2
	\forall
}
\newsavebox\forallBox
\newdimen\forallLineWidth
\newdimen\forallSep
\newcommand*{\fforallAux}[1]{%
	\sbox\forallBox{$\m@th#1\forall$}%
	\setlength{\forallLineWidth}{.06\wd\forallBox}%
	\setlength{\forallSep}{.09\wd\forallBox}%
	\tikz[
	inner sep=0pt,
	line cap=round,
	line width=\forallLineWidth,
	]
	\draw
	(0,0) node (A) {\copy\forallBox}
	(A.south) ++(-\forallSep-\forallLineWidth,.4\forallLineWidth)
	coordinate (A1)
	(A.north west) ++(-\forallSep,-\forallLineWidth)
	coordinate (A2)
	(A1) -- (A2)
	;%
}
\makeatother

\AtBeginDocument{%
  }

\setcopyright{acmlicensed}
\copyrightyear{2018}
\acmYear{2018}
\acmDOI{XXXXXXX.XXXXXXX}
\acmConference[RocqPL '26]{Rocq for Programming Languages}{June 03--05,
  2018}{Woodstock, NY}
\acmISBN{978-1-4503-XXXX-X/2018/06}

\begin{document}

\title{Recursive Mutexes in Separation Logic}

\author{Ke Du}
\email{kdu9@uic.edu}
\orcid{0009-0008-2465-1082}
\affiliation{%
  \institution{University of Illinois Chicago}
  \city{Chicago}
  \state{Illinois}
  \country{USA}
}

\author{William Mansky}
\email{mansky1@uic.edu}
\orcid{0000-0002-5351-895X}
\affiliation{%
  \institution{University of Illinois Chicago}
  \city{Chicago}
  \state{Illinois}
  \country{USA}
}

\author{Paolo G. Giarrusso}
\email{paolo@skylabs-ai.com}
\affiliation{%
  \institution{Skylabs AI}
  \city{}
  \state{}
  \country{}
}

\author{Gregory Malecha}
\email{gregory@skylabs-ai.com}
\orcid{0000-0003-3952-0807}
\affiliation{%
  \institution{Skylabs AI}
  \city{}
  \state{}
  \country{}
}

\renewcommand{\shortauthors}{Du et al.}

\newcommand{\mutexR}{\ensuremath{\mathsf{rmutex}}}

\begin{abstract}
Mutexes (i.e., locks) are well understood in separation logic, and can be specified in terms of either protecting an invariant or atomically changing the state of the lock. In this abstract, we develop the same styles of specifications for \emph{recursive} mutexes, a common variant of mutexes in object-oriented languages such as C++ and Java. A recursive mutex can be acquired any number of times by the same thread, and our specifications treat all acquires/releases uniformly, with clients only needing to determine whether they hold the mutex when accessing the lock invariant.
\end{abstract}



\maketitle

\section{Introduction}
By now, mechanized concurrent separation logic can prove correct programs in real languages using tools like VST~\cite{vst}, BRiCk~\cite{brick}, or RustBelt~\cite{rustbelt}.
To enable verifying more programs, specifying and verifying standard libraries is crucial.
Here, we report our progress on verifying C++'s concurrency library~\cite{ISO:2024:IIP} using BRiCk. While some utilities are standard, others have not been previously
analyzed in separation logic.
In particular, we address \emph{recursive mutexes}:
a thread owning a non-recursive mutex cannot acquire the same mutex again,
while a thread owning a recursive mutex can acquire it again, as shown in \cref{fig:rec-mutex}.
Crucially, methods \verb!update_a! and \verb!transfer! are both thread-safe, and
\verb!transfer! can reuse \verb!update_a! even if both acquire the same mutex!
With normal mutexes, \verb!transfer! couldn't call \verb!update_a!; typically,
\verb!transfer! could only call an alternative \verb!update_a_no_lock!.
To sum up, acquiring a recursive mutex is not just more convenient, but more
transparent and better encapsulated than using a normal mutex.


\begin{figure}[h]
  \small
\begin{verbatim}
class C {
  int balance_a; int balance_b;
  std::recursive_mutex mut;
public:
  void update_a(int x) {
    mut.lock(); balance_a += x; mut.unlock();
  }
  void update_b(int x) {
    mut.lock(); balance_b += x; mut.unlock();
  }

  void transfer(int x) {
    mut.lock(); update_a(x); update_b(-x); mut.unlock();
  }
};
\end{verbatim}
  \caption{An example using recursive mutexes.}
  \label{fig:rec-mutex}
\end{figure}

In this abstract we give a separation logic specification for recursive mutexes, and show that it enables idiomatic reasoning principles similar to those for regular mutexes.

\section{Specifying the recursive mutex}
The standard separation logic specification for a mutex/lock associates a lock $\ell$ with resources $P$ (the ``lock invariant''), represented as $\ell \mapsto_q \mathsf{lock}(P)$ (where $q$ is an ownership fraction of the lock). The lock functions are then specified by:
\begin{mathpar}
\{\ell \mapsto_q \mathsf{lock}(P)\}\ \texttt{lock}(\ell)\ \{\ell \mapsto_q \mathsf{lock}(P) \ast \mathsf{holds}(\ell) \ast P\}

\{\ell \mapsto_q \mathsf{lock}(P)\} \ast \mathsf{holds}(\ell) \ast P\}\ \texttt{unlock}(\ell)\ \{\ell \mapsto_q \mathsf{lock}(P)\}
\end{mathpar}
\noindent A thread that acquires the lock gains access to the protected resources, and gives them up upon release. Our goal is to state similar specifications for recursive mutexes. The key points are:
\begin{itemize}
\item A fractional handle $\ell \mapsto_q \mutexR\ \gamma\ P$ that is held by each thread that knows about the recursive mutex $\gamma$ with invariant $P$.
\item A predicate $\mathsf{holds}\ \gamma\ t\ n\ P$ indicating that thread $t$ has acquired mutex $\gamma$ with invariant $P$ $n$ times. When $n$ is 0, $t$ does not hold the mutex; when $n$ is greater than 0, it does, and can borrow $P$ from $\mathsf{holds}\ \gamma\ t\ n\ P$.
\end{itemize}
The main difference from the simple lock is that every thread that might interact with mutex $\gamma$, whether it currently holds it or not, has its own $\mathsf{holds}\ \gamma\ t\ n\ P$ assertion; $n$ is simply 0 for any thread that does not currently hold the mutex. 
The mutual exclusion property is captured by
$$
\mathsf{holds}\ \gamma\ t_1\ n_1\ P \ast \mathsf{holds}\ \gamma\ t_2\ n_2\ P \vdash n_1 = 0 \vee n_2 = 0
$$
Furthermore, it is vital that there is only one $\mathsf{holds}$ predicate per thread; otherwise, a thread could acquire the same lock twice and obtain $\mathsf{holds}\ \gamma\ t\ 1\ P \ast \mathsf{holds}\ \gamma\ t\ 1\ P$, which would permit borrowing two copies of $P$. 
Thus, the $\mathsf{holds}$ predicate must also satisfy $\mathsf{holds}\ \gamma\ t\ n_1\ P \ast \mathsf{holds}\ \gamma\ t\ n_2\ P \vdash \mathsf{False}$. 

From a programmer's perspective, the key feature of a recursive mutex is that we hold the protected resources after acquiring it, whether or not we held them before. Thus, our specs should let us conclude that we own the invariant after a call to \texttt{lock} without requiring case analysis on whether we already hold the lock. Our specifications allow this by saying that \texttt{lock} increases $n$ and \texttt{unlock} decreases it:
\begin{mathpar}
\{\ell \mapsto \mutexR\ \gamma\ q \ast \mathsf{holds}\ \gamma\ t\ n\ P\}\vspace{-.8em}\\ \vspace{-.8em}
\texttt{lock}(\ell)\\
\{\ell \mapsto \mutexR\ \gamma\ q \ast \mathsf{holds}\ \gamma\ t\ (n + 1)\ P\}

\{\ell \mapsto \mutexR\ \gamma\ q \ast \mathsf{holds}\ \gamma\ t\ (n + 1)\ P\}\vspace{-.8em}\\ \vspace{-.8em}
\texttt{unlock}(\ell)\\
\{\ell \mapsto \mutexR\ \gamma\ q \ast \mathsf{holds}\ \gamma\ t\ n\ P\}
\end{mathpar}
\noindent These specifications implicitly say that if $n$ was 0 before a \texttt{lock}, then afterwards $t$ gains ownership of $P$, and if $n$ is 0 after an \texttt{unlock}, then $t$ gave up ownership of $P$. However, these are not special cases in the specs---instead, threads may freely acquire and release $\ell$, incrementing and decrementing their counts $n$, and then extract $P$ from $\mathsf{holds}$ whenever they can show that $n$ is greater than 0. This allows simple proofs of code that use recursive mutexes idiomatically: whenever a thread needs access to $P$, it first acquires the mutex regardless of whether it already held it, and then releases it once finished, restoring the count to the value before the process.

The final piece is the ``borrowing'' rule that allows a thread to obtain the invariant $P$ if it holds the lock:
\begin{align*}
n > 0 \Rightarrow\ &\ell \mapsto \mutexR\ \gamma\ q \ast \mathsf{holds}\ \gamma\ t\ n\ P \dashv \vdash  \\&\ell \mapsto \mutexR\ \gamma\ q \ast P \ast \mathsf{held}\ \gamma\ t\ n
\end{align*}
\noindent A thread can exchange its positive $\mathsf{holds}$ predicate for ownership of $P$, plus a retainer $\mathsf{held}$ that tracks $n$ but does not give access to $P$.

 Essentially, while the standard lock acquire spec simultaneously acquires the lock and gives the thread ownership of $P$, our recursive mutex specs separate these operations into two steps, incrementing $n$ on acquire and then letting the thread obtain $P$ at any later point, as long as $n$ is greater than 0. Since $\mathsf{holds}$ is part of the precondition of \texttt{unlock}, the thread must unborrow $P$ before releasing the lock, whether or not the release actually sets $n$ to 0. 

\section{Fixing Arguments}
Lock invariants are usually written with existentially quantified arguments: for instance, a lock that protects the location $x$ may have the invariant $P \eqdef \exists v.\ x \mapsto v$. In this case, the borrowing rule's requirement that return $P$ to $\mathsf{holds}$ before acquiring or releasing the mutex effectively re-quantifies the variable, forcing the thread to lose information about resources it owns. We fix this by packing any such arguments into a tuple $a$, provided as an additional parameter to $\mathsf{holds}$: $\mathsf{holds}\ \gamma\ t\ n\ P\ a$ means that if $n > 0$, then we hold resources $P(a)$, and the argument $a$ will remain the same as long as $n$ remains positive. When $n = 0$, $P$ has returned to the invariant with quantified arguments, and $a$ is ignored. We use this approach to verify the example from \cref{fig:rec-mutex}.

The invariant for the example is \[P(a, b) \eqdef \verb!balance_a! \mapsto a \ast \verb!balance_b! \mapsto b\]

\noindent Our specification for \verb!update_a! is then:
\begin{mathpar}
\{\verb!mut! \mapsto \mutexR\ \gamma\ q \ast \mathsf{holds}\ \gamma\ t\ n\ P\ (a, b)\}\vspace{-.8em}\\ \vspace{-.8em}
\texttt{update\_a}(x)\\
\{\verb!mut! \mapsto \mutexR\ \gamma\ q \ast \mathsf{holds}\ \gamma\ t\ n\ P\ (a + x, b)\}
\end{mathpar}
\noindent and similarly for \verb!update_b!. When a thread calls \verb!update_a!, it gets back the lock with the same $n$ as before the call; if that $n$ is greater than 0 then we know that the new value of \verb!balance_a! is $a + x$, while otherwise we only know that the invariant $\exists a, b.\ P(a, b)$ has been preserved. Since \verb!transfer! begins by acquiring the mutex, we can show that the value of $\verb!balance_a! + \verb!balance_b!$ is preserved by the function.

\section{Discussion}
We have seen that the recursive mutex specs let us support the common pattern of functions that work equally well whether or not the caller holds the lock before calling, without losing any information about the invariant upon return. In our implementation, the specs are derived from more primitive \emph{logically atomic} specs and an associated invariant, but these details are entirely hidden from the client.

This work is part of our broader effort to verify the C++ standard library. We believe that this project will highlight more reasoning principles that are commonly used in mainstream software engineering.
Our current specifications include commonly used libraries such as \verb!std::atomic! and \verb!std::vector!.
We welcome collaboration on this project:
\begin{center}
\url{https://github.com/SkylabsAI/brick-libcpp}
\end{center}


\bibliographystyle{ACM-Reference-Format}
\bibliography{bibfile}

@inproceedings{iris,
  author    = {Ralf Jung and
               David Swasey and
               Filip Sieczkowski and
               Kasper Svendsen and
               Aaron Turon and
               Lars Birkedal and
               Derek Dreyer},
  title     = {Iris: Monoids and Invariants as an Orthogonal Basis for Concurrent
               Reasoning},
  booktitle = {Proceedings of the 42nd Annual {ACM} {SIGPLAN-SIGACT} Symposium on
               Principles of Programming Languages, {POPL} 2015, Mumbai, India, January
               15-17, 2015},
  pages     = {637--650},
  publisher = {{ACM}},
  year      = {2015},
  url       = {https://doi.org/10.1145/2676726.2676980},
  doi       = {10.1145/2676726.2676980},
  bibsource = {dblp computer science bibliography, https://dblp.org}
}

@book{vst,
  author    = {Andrew W. Appel and Robert Dockins and Aquinas Hobor and 
          Lennart Beringer and Josiah Dodds and Gordon Stewart and 
          Sandrine Blazy and Xavier Leroy},
  title     = {Program Logics for Certified Compilers},
  publisher = {Cambridge University Press},
  year      = {2014},
  url       = {http://www.cambridge.org/de/academic/subjects/computer-science/programming-languages-and-applied-logic/program-logics-certified-compilers?format=HB},
  isbn      = {978-1-10-704801-0},
  bibsource = {dblp computer science bibliography, https://dblp.org}
}

@String{pub-ISO                 = "International Organization for
                                  Standardization"}

@String{pub-ISO:adr             = "Geneva, Switzerland"}

@String{ack-nhfb = "Nelson H. F. Beebe,
                    University of Utah,
                    Department of Mathematics, 110 LCB,
                    155 S 1400 E RM 233,
                    Salt Lake City, UT 84112-0090, USA,
                    Tel: +1 801 581 5254,
                    FAX: +1 801 581 4148,
                    e-mail: \path|beebe@math.utah.edu|,
                            \path|beebe@acm.org|,
                            \path|beebe@computer.org| (Internet),
                    URL: \path|http://www.math.utah.edu/~beebe/|"}

@article{rustbelt,
    author = {Jung, Ralf and Jourdan, Jacques-Henri and Krebbers, Robbert and Dreyer, Derek},
    title = {{RustBelt}: Securing the Foundations of the Rust Programming Language},
    year = {2017},
    issue_date = {January 2018},
    publisher = {Association for Computing Machinery},
    address = {New York, NY, USA},
    volume = {2},
    number = {POPL},
    url = {https://doi.org/10.1145/3158154},
    doi = {10.1145/3158154},
    journal = {Proc. ACM Program. Lang.},
    month = {Dec},
    articleno = {66},
    numpages = {34}
    }

@Book{ISO:2024:IIP,
  author =       "{ISO}",
  title =        "{ISO\slash IEC 14882:2024 Programming languages ---
                 C++}",
  publisher =    pub-ISO,
  address =      pub-ISO:adr,
  edition =      "Seventh",
  pages =        "2104",
  month =        oct,
  year =         "2024",
  ISBN =         "????",
  ISBN-13 =      "????",
  LCCN =         "????",
  bibdate =      "Wed Apr 30 14:48:53 2025",
  bibsource =    "https://www.math.utah.edu/pub/tex/bib/isostd.bib",
  URL =          "https://www.iso.org/standard/83626.html",
  acknowledgement = ack-nhfb,
}

@misc{brick,
  title = "The {BRiCk} Project",
  authors = "Gregory Malecha, Paolo Giarrusso, David Swasey, Hai Dang, František Farka, Abhishek Anand, Hai Dang",
  howpublished = {\url{https://github.com/SkylabsAI/BRiCk}},
  year = {2025}
}

\end{document}